\documentclass[12pt]{article}
\usepackage{psfig,epsfig}
\begin{document} 
\def\cc#1{$^{[#1]}$}
\def\bcc{\begin{center}} \def\ecc{\end{center}}
\def\beq{\begin{equation}} \def\eeq{\end{equation}}
\def\bea{\begin{eqnarray}} \def\eea{\end{eqnarray}}
\def\beqa{\begin{eqnarray}}  \def\eeqa{\end{eqnarray}}
\def\nl{\null{}} \def\la{\langle} \def\ra{\rangle}
\def\btm{\begin{itemize}} \def\etm{\end{itemize}}
\def\nnu{\nonumber} \def\vs{\vskip}
\def\cl{\centerline}  \def\d{\rm d}
\def\f{\left}  \def\g{\right} \def\n{\noindent} \def\pl{\partial}
\def\rf{\varphi}  \def\pt{p_{\rm t}} \def\pj{{\rm j}}
\setcounter{page}{1}

\centerline{\Large On the Erraticity in Random-Cascading\ $\alpha$
  Model\footnote{Supported in part by the NSFC under
  Grant No 19975021}}

\medskip
\centerline{ZHOU Yi-Fei \hskip0.7cm LIU Qin \hskip0.7cm 
TANG Ying\hskip0.7cm CHENG Chun\footnote{The authors are students of the 
State-Level Personnel Training Base for Research and Teaching in Fundamental 
Sciences (Physics).}}

\medskip

\centerline{\small (Institute of Particle Physics, Huazhong Normal University 
\hskip0.25cm Wuhan\hskip0.25cm  430079) }

\vskip2cm
\centerline{\large Abstract}

\medskip
{\small The erraticity in the random-cascading $\alpha$ model is revisited.
It is found that in contrary to the previous expectation, even in the pure 
single-$\alpha$ random-cascading model without putting in any particle 
there exists erraticity behavior and the corresponding entropy indices 
do not vanish. This means that the dynamical fluctuations in a pure 
single-$\alpha$ model already fluctuate event-by-event. Models with 
multiple-$\alpha$ strengthen this fluctuation. Taking double-$\alpha$ 
model as example, the variation of the event-space fluctuation strength 
with the mixing ratio of the two $\alpha$'s is studied in some detail.
The influence of particle number on the results when particles are putted 
into the system is also investigated.}

\vskip3cm
{\small Keywords:\ \  Dynamical fluctuation  
   \hskip0.5cm random cascading $\alpha$ model \hskip0.5cm erraticity 

 \hskip2.2cm entropy index }  
\newpage


Since the observation of large local fluctuations in a high multiplicity
nucleus-nucleus collision event induced by cosmic ray in 1983\cc{1}, similar 
phenomenon has also been observed in the accelerator experiments\cc{2}.
Since then, extensive attentions are paid to the non-linear phenomena 
in high energy collisions\cc{3}.

Due to the finiteness of particle number, the particle distribution fluctuates
with respect to the probability distribution obtained from theoretical
calculation. This is the so called statistical fluctuations. For a 
long time the popular idea was that the probability distribution obtained 
from theoretical calculation is smooth and the actual particle distribution
fluctuates statistically on top of the smooth probability distribution.
However, is it really the case? 
Is it possible that there are non-linear dynamics which causes 
the dynamical probability distribution itself to be non-smooth? This question 
is still to be answered. Although a systematic quantum non-linear theory 
has not been constructed yet and a definite reply to this quesition 
is still impossible by now, non-linear phenomena have already been 
observed in high energy experiments\cc{4--6} as stated above.

In studying the non-linear dynamical fluctuations, besides the factorial 
moments averaged over event sample, the importance of the fluctuations of 
single-event moments in the event samples has also been noticed\cc{7}.
It is shown that these fluctuations are associated with the system's chaotic
behavior.  A parameter $\mu$, called entropy index, is introduced to describe 
the chaotic behavior. An entropy index greater than zero is the signal of
chaos. This method is called ``erraticity analysis''.

The random-cascading $\alpha$ model is a simple model which can simulate 
the non-linear dynamical fluctuations. It has been used in Ref.[8--11] to 
study the dynamical fluctuations in event space, i.e. erraticity. Since the 
value of parameter $\alpha$ determines the strength of dynamical fluctuations, 
it was argued in Ref.[9] that there is no erraticity in the model with a
single parameter $\alpha$, and therefore a model with Gaussian-distributed 
$\alpha$ was proposed instead.

In this letter the erraticity of $\alpha$ model is revisited. It is found
that there is erraticity already in a pure single-$\alpha$ model. 
The corresponding entropy index is small but does not vanish. 
It increases with the increasing of $\alpha$. 
The mixing of different $\alpha$'s makes the entropy index increases further. 
The dependence of entropy index on the mixing ratio is studied. 
It is shown through putting in particles that in all cases
the entropy index calculated by particle number is bigger than that in pure 
$\alpha$ model and tends to the latter only when the particle number tends 
to infinity. This reflects the effect of statistical fluctuations.


Let us first briefly recall the definition of erraticity and
entropy index\cc{7}.\

The single-event factorial moments $F_{q}^{(e)}$  and probabilitic 
moments $C_q^{(e)}$ are defined as 
\beq  
{F_{q}^{(e)}}={{\frac{1}{M}\sum\limits_{m=1}^{M}n_{m}(n_{m}-1) \cdots 
 (n_{m}-q+1)}\over{{\f(\frac{1}{M}\sum\limits_{m=1}^{M}n_{m}\g)}^{q}}},
\end{equation}
\beq  
C_q^{(e)} = M^{q-1}\sum_{m=1}^M \f( p_m^{(e)}\g)^q,
\eeq
respectively, where a phase space region is divided into $M$ equivalent 
sub-regions, $p_m$ and $n_m$ are correspondingly the probability and particle 
numbers in the $m$th sub-region.

The fluctuation of moments ${F_{q}^{(e)}}$ and $C_q^{(e)}$ in the event
space can be characterized by their event-space moments 
\beq  
C_{p,q}^{(F)} = \la F_q^{(e)p}\ra/ \la F_q^{(e)}\ra^p, \qquad
C_{p,q}^{(C)} = \la C_q^{(e)p}\ra/ \la C_q^{(e)}\ra^p.
\eeq
If $C_{p,q}^{(F)}$ or $C_{p,q}^{(C)}$ has power law  behavior\
\beq   
C_{p,q}^{(F)} \propto M^{\psi_q^{(F)}(p)}, \qquad
C_{p,q}^{(C)}  \propto M^{\psi_q^{(C)}(p)}, 
\eeq
when $M$ is big then we say that there is erraticity in the system. It can 
be characterized quantitatively by the entropy index 
\beq  
 \mu_q =\f. \frac{\d}{\d p}\psi_q(p)\g|_{p=1}.
\eeq

Another method for calculating entropy index is to express $C_{p,q}$ as 
\beq  
C_{p,q} = \la \Phi_q^{(e)p}\ra,
\eeq
where
\beq  
\Phi_q^{(e)} = F_q^{(e)} / \la  F_q^{(e)} \ra, \qquad
\Phi_q^{(e)} = C_q^{(e)} / \la  C_q^{(e)} \ra.
\eeq
Define
\beq  
\Sigma_q = \la \Phi_q^{(e)} \ln \Phi_q^{(e)} \ra,
\eeq
then in the scaling region, i.e. in the region where $\Sigma_q$ depends on 
$\ln M$ linearly, we can write
\beq  
\mu_q = \frac{\partial \Sigma_q}{\partial \ln M}.
\eeq


In the following we will study the erraticity in the random-cascading 
$\alpha$ model\cc{12}. In this model the $M$ partition of the phase 
space region $\Delta$ is realized in $\nu$ steps. At the first step, it is 
divided into two equal parts; at the second step, each part is further divided 
into two parts, \dots, and so on. The steps are repeated until 
$M=\Delta Y / \delta y= 2^{\nu}$. In this step-by-step partition how particles 
are distributed between the two parts of a given phase space cell is determined 
by the value of independent random variable ${\omega}_{\nu{j_{\nu}}}$, where 
$j_{\nu}$is the position of the sub-cell ($(1{\leq}{j_{\nu}}{\leq}{2^{\nu}})$),
$\nu$ is the number of steps. The value of the random variable ${\omega}_{\nu{j_{\nu}}}$
is given by:
\beq  
\omega_{\nu,2j-1}=\frac{1+\alpha r}{2},\qquad
\omega_{\nu,2j}=\frac{1-\alpha r}{2},\qquad
j=1,...,{2^{\nu-1}}
\eeq
where $r$ is a random number distributed uniformly in the interval [-1,1],
$\alpha$ is a positive number less than or equal to unity, which determines the 
value-region of the random variable $\omega$, and thus describes the 
strength of dynamical fluctuations in the model. After $\nu$ steps, the 
probability in the $m$th sub-cell ($m=1,\dots,M$) is
\beq  
{p_{m}}={\omega_{1j_{1}}}{\omega_{2j_{2}}}\cdots {\omega_{\nu j_{\nu}}}.
\eeq
Using these probabilities the probability moment $C_q^{(e)}$ in each event 
is calculated according to (2), and the entropy index $\mu_q$ is obtained 
through Eq's.(2)-(9).

Besides the single-$\alpha$ model described above, the double-$\alpha$-mixing
model will also be used later. It is constructed by letting $r_1\cal N$ 
events among the total $\cal N$ events have $\alpha=\alpha_1$, and the 
remaining $(1-r_1)\cal N$ events have $\alpha=\alpha_2$.

In order to study the influence of limited particle number, we put a certain
number $N$ of particles into the $M$ subcells according to the Bernauli 
distribution  
\beq  
B(n_1,\dots,n_M|p_1,\dots,p_M) =\frac{N!}{n_1!\cdots n_M!}
      p_1^{n_1}\cdots p_M^{n_M}.
\eeq
The particle numbers $n_m\ (m=1,\dots,M)$ in every window are thus
obtained and the event factorial moments $F_q^{(\rm e)}$ can be calculated.
The entropy index $\mu_q$ is then obtained through Eq's. (2)--(9). 


Eq.'s (5) and (9) are two equivalent methods for getting
entropy index. We now discuss which one is more suitable for our purpose.

What we are interested in is the case when the partition number $M$ of
phase space is very big. When the number of particles is small and that
of sub-cells is big, it is probable that the particle number in every window 
is 0 or 1. It can be seen from Eq.(1) that this will result in 
$F_{q}^{(e)}=0$ for all $q>1$, and then lead to $\Phi_{q}^{(e)}$ and
$\sum_q^{(e)}$ both vanish. Therefore, the  second method fails in this
case. So we choose the first method to get $\mu_{q}$.

When using the first method, it can be seen from Eq.(5) that $\mu_{q}$ is 
obtained through partially differentiating $\psi_q^{(\rm e)}$ with respect
to $p$ near $p=1$. Then how to choose the range 

\leftskip 3in
\noindent  of $p$ value near $p=1$ to 
fit the partial derivative is also a question which we must answer.
A too small range will lead to a wrong result due to the limitation of 
computer precision. To make the choise we show in Fig.1 the 
$\psi$-$p$ plot. It can be seen from the figure that within the range 
$0.5 \leq p\leq 2$ $\psi$ versus $p$ is apprximately linear.
Therefore, we take it as the right range for fitting the partial derivative
$\pl\psi/\pl p$.

     \begin{center}
     \begin{picture}(60,60)
     \put(-186,86)
     {
     {\epsfig{file=fig1.epsi,width=170pt,height=170pt}}
     }
     \end{picture}
     \end{center}
\vskip -2.0cm
\medskip
\leftskip 0in

\vskip -1.5cm
\hskip 1.0cm Fig.1 $\psi \sim p$ relation

\vskip0.5cm

In real calculation when $\alpha$ is very small (e.g. $\alpha$=0.1)
the value of single event probability moment $C_q^{(e)}$ is close to 
zero. This results in big errors in numerical calculation.
In order to solve this problem, we multiply the value of single event 
probability moments by a large number, e.g. 1000. This extra factor will
appear in both the denominator and the numerator in the formula for
calculating $C_{p,q}$, Eq. (3), and will not effect the final results. 


The calculating results of single $\alpha$ model show that there exists 
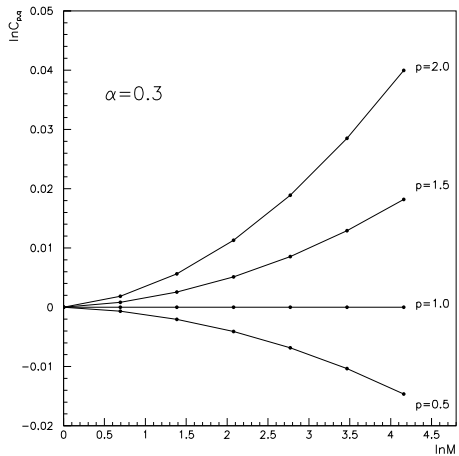
erraticity. As example, in Fig.2 is shown the variation of $C_{p,q}$ with $M$
for $\alpha$=0.3. It shows a typical characteristic of erraticity.
The entropy indices $\mu_2$ in this model for various values of
$\alpha$ are plotted at the right side of Fig.3. It can be seen that
$\mu_2>0$, indicating the existence of erraticity. The value of entropy 
index $\mu_2$ increases with the increasing of $\alpha$.

     \begin{center}
     \begin{picture}(60,60)
     \put(-165,-100)
     {
     {\epsfig{file=fig2.epsi,width=170pt,height=170pt}}
     }
     \end{picture}
     
\end{center}

     \begin{center}
     \begin{picture}(60,60)
     \put(45,-30)
     {
     {\epsfig{file=fig3.epsi,width=170pt,height=170pt}}
     }
     \end{picture}
     \end{center}

\vskip0.6cm
\hskip1cm
Fig.2\ The variation of $C_{p,q} \sim M$ 
\hskip2cm
Fig.3 \ Entropy index for single 

\hskip1cm
for $\alpha=0.3$ \hskip6cm
$\alpha$ model

 On the left of Fig. 3 is shown the variation of entropy index with the
number of particles when particles are putted into the model.
It can be seen that when the number of particles is small the
corresponding entropy index is much bigger than that of the pure
$\alpha$ model without particle.  The entropy index decreases with the 
increasing of the number of particles, tends to the value of the pure 
$\alpha$ model when the number of particles tends to infinity.

The reason why erraticity exists also in the single-$\alpha$ model is that,
in the expression Eq. (10) for the elementary probability in the model, 
there is a random number $r$, causing the window probability to fluctuate 
inspite of the fact that the values of $\alpha$ are the same in different 
events. 

It is clear that if the $\alpha$'s are different in different 
events, the fluctuation will be strengthened. We take the double-$\alpha$ 
model as example to study this situation.

The two $\alpha$'s are taken to be $\alpha_1=0.2, \alpha_2=0.8$.
There are in total $\cal N$ events amoung which
$r_1{\cal N}$ events  have $\alpha_1=0.2$ and the remaining 
$(1-r_1){\cal N}$  events have $\alpha_1=0.8$.

In Fig.4 are shown the log$_{10}\mu$ versus the number of particles $n$
for $r_1= 0,0.1,0.3,0.5,0.7,$ $0.9,0.95,0.98,1$, repectively.
It can be seen that the entropy index $\mu$ firstly increases with the 
increasing of $r_1$ starting from $r_1=0$, but when $r_1$ is big (near
to unity) it turns to 

\leftskip 3in
\noindent
decreasing. The reason is that $r_1=0$ and $r_1=1$
correspond to single-$\alpha$ model with $\alpha=0.2$ and 0.8 respectively.
The entropy indices for these two cases are in general smaller than that of
the cases with the two $\alpha$ values mixed. The only exception is when
only a very little fraction ( $< 0.1$) of big $\alpha$ (0.8) is mixed in.
In this case, the double-$\alpha$-mixing model will have a smaller entropy 
index than the single-$\alpha$ model with $\alpha=0.8$.

     \begin{center}
     \begin{picture}(60,60)
     \put(-194,95)
     {
     {\epsfig{file=fig4.epsi,width=170pt,height=170pt}}
     }
     \end{picture}
     \end{center}

\leftskip 0in
\vskip-4.0cm Fig.4 \ Entropy index versus 

\hskip0.5cm multiplicity for different 

\hskip0.5cm double-$\alpha$ model 

The variation of entropy index in double-$\alpha$-mixing model with the
mixing ratio $r_1$ is plotted in Fig.5 for 
$\alpha_1,\alpha_2=0.1,0.9;0.2,0.8;0.3,0.7;0.4,0.6$, respectively.
The increasing of $\mu$ with $r_1$ when the latter is not very big
and the decreasing of $\mu$ when $r_1$ near to unity can be seen in 
all the cases.

In order to show the fine structure near $r_1=1$ the curves for 
$r_1,r_2=$ 0.1,0.9 (case-A) and 0.4,0.6 (case-B) are magnified and shown in
the small figure inside Fig.5. The crossing of these two curves is
understandable. \ This is because \ 
$\alpha_1^{(\rm A)} < \alpha_1^{(\rm B)}$
\ while \ $\alpha_2^{(\rm A)} > \alpha_2^{(\rm B)}$. 

     \begin{center}
     \begin{picture}(60,60)
     \put(-191,-109)
     {
     {\epsfig{file=fig5.epsi,width=170pt,height=170pt}}
     }
     \end{picture}
     \end{center}

\vskip3.2cm
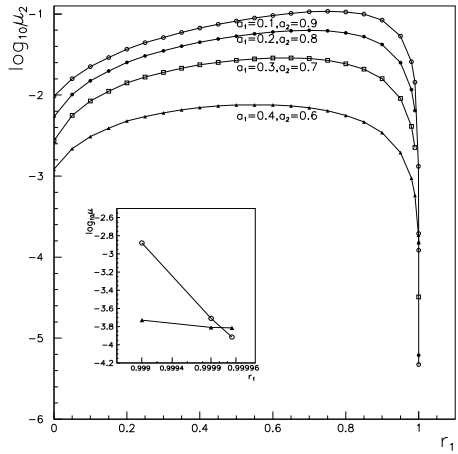
\ \   \noindent{Fig.5\ Entropy index in double-$\alpha$}

\vskip-0.2cm
model as function of mixing ratio

\leftskip3in
\vskip-7.5cm
\noindent Therefore, when $r_1$
is small (1-$r_1$ is big) $\mu^{(\rm A)} > \mu^{(\rm B)}$, but when
$r_1$ is big (1-$r_1$ is small) $\mu^{(\rm B)} > \mu^{(\rm A)}$.

In summary, it is found in this letter that even in the pure 
single-$\alpha$ random-cascading model without putting in any particle 
there exists erraticity behavior and the corresponding entropy indices 
do not vanish. This observation is in contrary to the previous expectation.
It means that the dynamical fluctuations in a pure 
single-$\alpha$ model already fluctuate event-by-event. 

\leftskip0in
This fluctuation is strengthened when different values of $\alpha$ are
mixed in the model.  Taking double-$\alpha$-mixing 
model as example, the variation of the entropy index
with the mixing ratio of the two $\alpha$'s is studied.
It is found that the entropy index $\mu$ firstly increases with the 
increasing of $r_1$ and then decreases. The maximum is located near
$r_1=1$ when $\alpha_1<\alpha_2$.

The influence of particle number on the results when particles are putted 
into the system is also investigated. In all cases
the entropy index calculated by particle number is bigger than that in pure 
$\alpha$ model and tends to the latter only when the particle number tends 
to infinity. 

\vskip0.8cm
\noindent Acknowledgement

The authors show sincere thanks to the patient direction given by Liu
 Lianshou and the warm help given by Liu Fuming.


\newpage

\begin{thebibliography}{9}
\itemsep=-1mm

\def\Journal#1#2#3#4{{#1} {\bf #2} (#3) #4}
\def\NCA{\em Nuovo Cimento} \def\NIM{\em Nucl. Instrum. Methods}
\def\NIMA{{\em Nucl. Instrum. Methods} A} \def\NPB{{\em Nucl. Phys.} B}
\def\PLB{{\em Phys. Lett.}  B} \def\PRL{\em Phys. Rev. Lett.}
\def\PRD{{\em Phys. Rev.} D} \def\ZPC{{\em Z. Phys.} C}
\def\PRE{{\em Phys. Rev.} E} \def\PRC{{\em Phys. Rev.} C}

\bibitem{[1]} 
 T. H. Burnett et al., \Journal{\PRL}{50}{1983}{2062}.

\bibitem{[2]} 
M. Adamus et al. (NA22), \Journal{\PLB}{185}{200}{1987}.

\bibitem{[3]} 
E.A. De Wolf, I.M. Dremin and W. Kittel, {\em Phys. Rep.}
 {\bf 270} (1996) 1.

\bibitem{[4]} 
N.M. Agababyan et al. (NA22), {\em Phys. Lett.} {\bf B382} (1996) 305;
{\em ibid} {\bf B431} (1998) 451.

\bibitem{[5]} 
S. Wang, Z. Wang and C. Wu, \Journal{\PLB}{410}{323}{1997}.

\bibitem{[6]} 
Liu Feng, Liu Fuming and Liu Lianshou, \Journal{\PRD}{59}{114020}{1999}.

\bibitem{[7]} 
Z. Cao and R. Hwa, \Journal{\PRL}{75}{1995}{1268};
Z. Cao and R. Hwa, \Journal{\PRD}{54}{1996}{6674};
Z. Cao and R. Hwa, \Journal{\PRE}{56}{1997}{326}.

\bibitem{[8]} 
Wang S.S. and Wang Z.M., \Journal{\PRD}{57}{1998}{3036}.

\bibitem{[9]} 
Fu Jinghua, Wu Yuanfang, Liu Lianshou.
 {\em HEP \& NP} {\bf 23} (1999) 673.

\bibitem{[10]} 
Fu Jinghua, Wu Yuanfang and Liu Lianshou,
\Journal{\PLB}{472}{2000}{161}.

\bibitem{[11]} 
Liu Lianshou, Fu Jinghua and Wu Yuanfang,
{\em Science in China} {\bf A30} (2000) 432.

\bibitem{[12]} 
Wu Yuanfang, Zhang Kunshi and Liu Lianshou, {\em
Chinese Science Bulletin} {\bf 36} (1991) 1077.

\end{thebibliography}
\end{document}